# Rich structural polymorphism of monolayer $C_{60}$ from cluster rotation

Xueao Li[#], Fan Zhang[#], Xuefei Wang, Weiwei Gao[*], and Jijun Zhao

*Key Laboratory of Material Modification by Laser, Ion and Electron Beams (Dalian University of Technology), Ministry of Education, Dalian 116024, China*

# X.L. and F.Z. contributed equally to this work.

**ABSTRACT**

The recent experimental fabrication of monolayer and few-layer $C_{60}$ polymers paves the way for synthesizing two-dimensional cluster-assembled materials. Compared to atoms with the SO(3) symmetry, clusters as superatoms (e.g., $C_{60}$) have an additional rotational degree of freedom, greatly enriching the phase spaces of superatom-assembled materials. Using first-principles calculations, we find the energy barriers of cluster rotation in quasi-tetragonal monolayer $C_{60}$ structures are rather low (about 10 meV/atom). The small rotational energy barriers lead to a series of tetragonal $C_{60}$ polymorphs with energies that are close to the experimental quasi-tetragonal (expt-qT) phase. Similarly, several dynamically stable quasi-hexagonal monolayer $C_{60}$ structures are found to have energies within 7 meV/atom above the experimental quasi-hexagonal phase. Our calculations demonstrate photo-excited electron-hole pairs and electrostatic doping of electrons can effectively modulate the relative energies of quasi-tetragonal $C_{60}$ polymorphs. Particularly, the unstable monolayer expt-qT phase becomes dynamically stable when it is electrostatically doped with electrons. In contrast, the relative energies between different quasi-hexagonal polymorphs are insensitive to electrostatic doping of electrons.

[*] Corresponding Authors. Email: weiweigao@dlut.edu.cn

**INTRODUCTION**

Conventionally, atoms are considered as fundamental building blocks of materials. However, analyzing or designing materials by examining the chemical interactions between atoms as well as the numerous possible combinations of elements is not always the most convenient route. Many materials can be considered as assemblies of "superatoms"[1-4], which are structurally stable entities (or clusters) consisting of strongly bounded atoms and show chemical behavior similar to an individual atom[5-9]. The concept of superatom not only provides a new perspective on cluster-assembled materials, but also opens up new avenues for exploring the building blocks of novel functional materials[3, 10].

Among candidates of superatoms, $C_{60}$ buckminsterfullerene has been widely studied due to several desirable properties[11-13]. First of all, $C_{60}$ is one of the few nanoclusters that can be mass-produced in high quality[14]. It exhibits exceptional stability under high temperature and a wide range of chemical environments, making it a promising building block of cluster-assembled materials[15]. Bulk $C_{60}$-based crystal has a rich temperature-pressure phase diagram[16]. Under ambient conditions, $C_{60}$ solid is a face-centered cubic (f.c.c.) solid bounded by van der Waals (vdW) interaction[17]. Under suitable pressure and temperature, $C_{60}$ solids undergo a series of phase transitions to different carbon allotropes, such as quasi-low-dimensional polymers with tetragonal or hexagonal cells[18-19], disordered states[20], simple cubic phase[20-22], and many more[23-24]. Theoretically, several two-dimensional (2D) $C_{60}$ polymers have also been proposed[25-28], while synthesis of these monolayer $C_{60}$ crystals has proven challenging. Recently, researchers successfully synthesized two-dimensional polymeric $C_{60}$ single crystals, namely, quasi-hexagonal (qH) phase and quasi-tetragonal (qT) phase $C_{60}$, by removing $Mg^{2+}$ ions from magnesium fulleride $Mg_4C_{60}$ or $Mg_2C_{60}$, respectively[29-30]. However, only monolayer qH phase was achieved, while the qT phase was synthesized as a few-layer material.

In contrast to an individual atom, which is invariant under rotations about its center, a cluster or superatom has a finite point group symmetry and hence an additional rotational degree of freedom, which, in turn, extends the phase space and affects the physical properties of cluster-based materials. $C_{60}$ clusters, for example, rotate freely in the f.c.c. phase at ambient temperature but are fixed in an ordered orientation in the simple cubic phase between 100 and 260 K[22]. Below 85 K, the orientations of $C_{60}$ molecules are frozen disorderly in a glassy state[20, 22, 31]. Quasi-one-dimensional (1D) and 2D $C_{60}$ polymeric phases have also been proposed to show various $C_{60}$ orientations and physical properties[32-33]. In this work, we investigated the energies and structural properties of monolayer quasi-hexagonal and quasi-tetragonal $C_{60}$ polymorphs that are related by rotating $C_{60}$ cages. Similar to bulk phases of $C_{60}$-based crystals bonded with vdW interaction, $C_{60}$ fullerenes in the monolayer polymeric quasi-tetragonal phases have low rotating energy barriers, suggesting the coexistence of different polymorphs or even orientationally disordered states. Electrostatic doping of electrons and optically excited electron-hole pairs effectively switch the relative energies and induce phase transitions between quasi-tetragonal polymorphs. Different from quasi-tetragonal phases, monolayer quasi-hexagonal phases are insensitive to electrostatic doping and have more rigid frameworks where $C_{60}$ fullerenes are difficult to rotate freely.

## RESULTS AND DISCUSSIONS

As shown in Fig. 1 (a), the experimental quasi-tetragonal phase (abbreviated as "expt-qT" thereafter) has mixed types of inter-cluster bonds, including [2+2] cycloaddition bonds connecting $C_{60}$ clusters along the *b*-direction and single C-C bonds connecting $C_{60}$ along the *a*-direction[29]. Fully structural optimization of the monolayer expt-qT phase shows that the C-C single bonds along the *a*-direction are broken, resulting in periodically arranged 1D chain-like (abbreviated as "1D-chain") $C_{60}$ polymers with [2+2] cycloaddition bonds along the *b*-direction. Similarly, the experiments did not report a viable synthesis of monolayer expt-qT $C_{60}$ [29] and recent computational work also suggests that monolayer expt-qT structure is unstable[34].

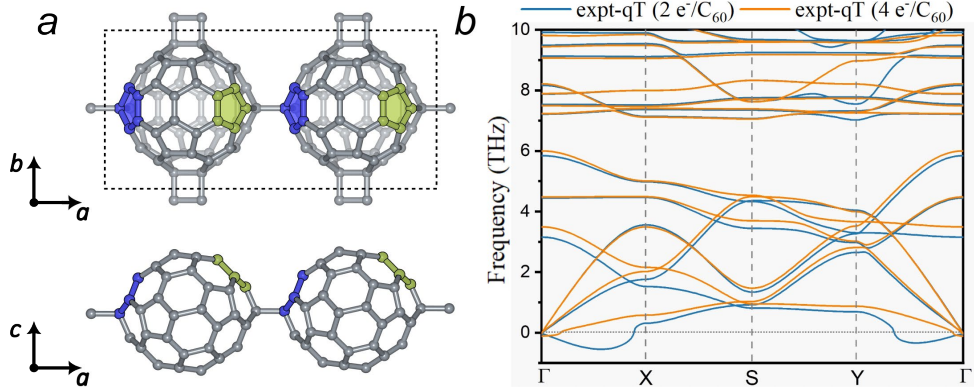

Fig. 1 (a) The top and side views of the experimentally synthesized few-layer quasi-tetragonal (expt-qT) phase. Pentagons are highlighted in colors to show the orientation of $C_{60}$ clusters. (b) Phonon dispersions of monolayer expt-qT phase with 2 and 4 electrons per $C_{60}$ doping. Undoped expt-qT structure decompose to 1D-chain structures and its phonon spectra is not shown. For clarity, only phonon modes with frequencies lower than 10 THz are shown.

The experimentally synthesized few-layer quasi-tetragonal $C_{60}$ polymers are prepared from $Mg_2C_{60}$, which crystallizes in a monoclinic space group I2/m. Given that $C_{60}$ is an electron acceptor in fullerides[35], one might expect that the expt-qT structure becomes stable upon electrostatically doping of electrons. To put this theory to the test, we found that electrostatically doping the expt-qT phase with 1.25 electrons per $C_{60}$ keeps the C-C single bonds intact during structural relaxation. The stability of the expt-qT polymer can be further confirmed by checking the phonon dispersion under doping conditions. As shown in Fig. 1 (b), the monolayer expt-qT structure has imaginary phonon modes in a small region around Γ point when it is doped with 2 electrons/$C_{60}$. As the doping electron concentration further increases to 4 electrons/$C_{60}$, the imaginary phonon modes disappear.

To gain deeper understanding of the energy landscapes of quasi-tetragonal monolayer $C_{60}$ polymers, we compared expt-qT with 1D-chain as mentioned above and another previously found tetragonal $C_{60}$ polymer (named as qT-E phase for brevity) [36-38], which has [2+2] cycloaddition bonds bridging $C_{60}$ clusters along both $a$- and $b$-directions. The energies of qT-E, expt-qT, and 1D-chain structures with respect to the lattice constant $a$ for a series of electron doping concentrations are presented in Fig. 2 (a-c), where the energy of the 1D-chain structure is set as the reference. For electron concentrations ranging from 0 to 6 e⁻/$C_{60}$, the energy of 1D-chain is lower than those of expt-qT and qT-E phases. Under the undoped condition, the energy curve of the expt-qT superimposes with that of 1D-chain and does not show a local minimum for lattice constant $a$ less than 9.6 Å. This suggests the expt-qT phase directly decomposes to 1D-chain without electron doping. As the electron concentration increases to 2 $e^-/C_{60}$, a local minimum (or saddle point) appears in the energy curve of the expt-qT phase. The local minimum has higher energy than the qT-E phase by about 7 meV/atom. As the electron concentration increases to more than 6 $e^-/C_{60}$, the energy of expt-qT phase becomes lower than that of the qT-E phase. The reversal of relative energy ordering between the qT-E and expt-qT structures suggests that electrostatic doping of electrons is an effective approach to induce a structural transition between these two polymorphs.

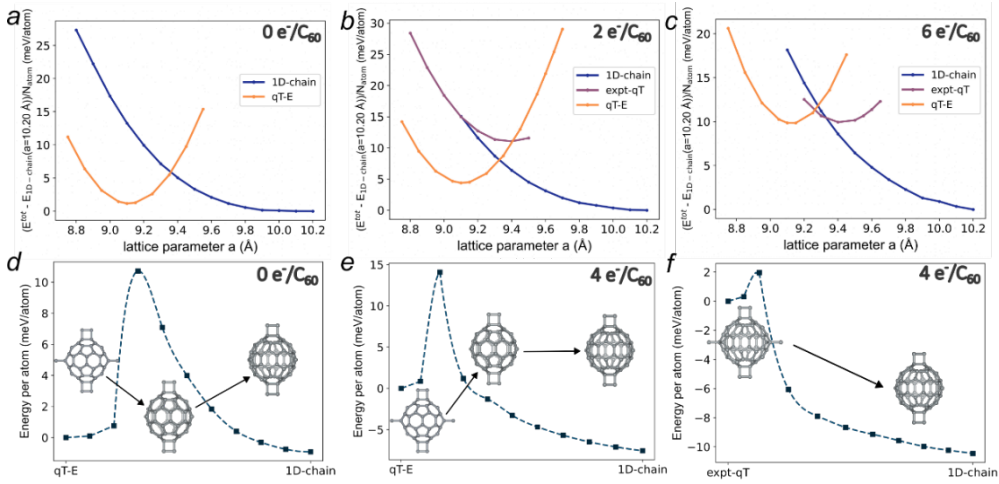

Fig. 2 (a-c) The energies of expt-qT, qT-E, and 1D-chain versus the lattice parameter $a$ with electrostatically doping electrons of 0, 2, and 6 electrons per $C_{60}$, respectively. (d-e) The changes in energy along the transition path from qT-E to 1D-chain structure when the material is doped with 0 and 4 electrons per $C_{60}$, respectively. (f) The changes in energy along the transition path from expt-qT to 1D-chain with an electron doping concentration of 4 per $C_{60}$.

We calculated the energy barriers for structure transitions between qT-E, qT-expt, and 1D-chain structures using the nudged-elastic band (NEB) method[39-40]. The transitions between these three structures mainly involve the rotation of $C_{60}$ clusters around the $b$-axis. Figure 2 (d) and (e) illustrate the energy changes along the transition path from qTP-E to 1D-chain structure under undoped and doped conditions, respectively, with schematic plots of initial, intermediate, and final structures. During the transition, the inter-cluster bonds along the $a$-direction of qTP-E are broken in the first few steps, leading to a drastic energy increase, which is then followed by a gradual transition to 1D-chain structure. The energy barrier between qT-E and 1D-chain in the undoped case is approximately 11 meV/atom, which increases to 14 meV per atom upon doping 4 $e^-/C_{60}$. Similarly, Fig. 2 (f) shows the energy changes along the transition from expt-qT to the 1D-chain structure under 4 $e^-/C_{60}$ doping condition. The energy barrier between expt-qT and 1D-chain is only 2 meV/atom. Such a small transition barrier is even lower than the rotation energy barrier of 6 meV/atom in the f.c.c. $C_{60}$ solid with vdW inter-cluster interaction (as shown in Fig. S1 and S2 in the Supplementary Information). In other words, monolayer expt-qT structure can only exist at low-temperature even it is doped with electrons. At the ambient condition, thermal fluctuation may drive a structural transition from expt-qT to 1D-chain or qT-E structure.

The small rotating energy barrier of $C_{60}$ in the quasi-tetragonal structures suggests the possibility of alternative polymorphs composed of fullerene clusters oriented in different directions. We examined three possible orientations of $C_{60}$ clusters, labeled as I, II, and III, which appear in expt-qT and qT-E structures. Note that I and III are related by a mirror-reflection symmetry operation with respect to the *bc*-plane, as shown in Fig. 3(a). A collection of $C_{60}$ polymeric quasi-tetragonal (qT) allotropes, named as qT-A, B, C, D, E, F, and G structures, are illustrated in Fig. 3 (b). We optimized the structures of these quasi-tetragonal $C_{60}$ polymers and compared their energies under undoped and electrostatic doping conditions. The considered range of carrier doping concentration is within $4.8\times10^{14}$ cm$^{-2}$, which is achievable in 2D materials with gate-induced electrostatic doping[41]. As shown in Fig. 3(c), among these 2D qT polymorphs, the qT-E structure remains the ground state for doping-carrier concentration in range of 1 to 4 $e^-/C_{60}$. The qT-A, B, C, and D structures can exist in a limited range of electron doping concentrations and have slightly higher energy compared to the expt-qT phase. Interestingly, with doping-carrier concentration in range of 1 to 3 $e^-/C_{60}$, both qT-F and qT-G structures have lower energies than the expt-qT phase. Overall speaking, the energy differences between these tetragonal structures are small (less than 12 meV/atom), indicating the coexistence of several qT phases or even an orientational glass state in a 2D quasi-tetragonal C$_{60}$ polymer.

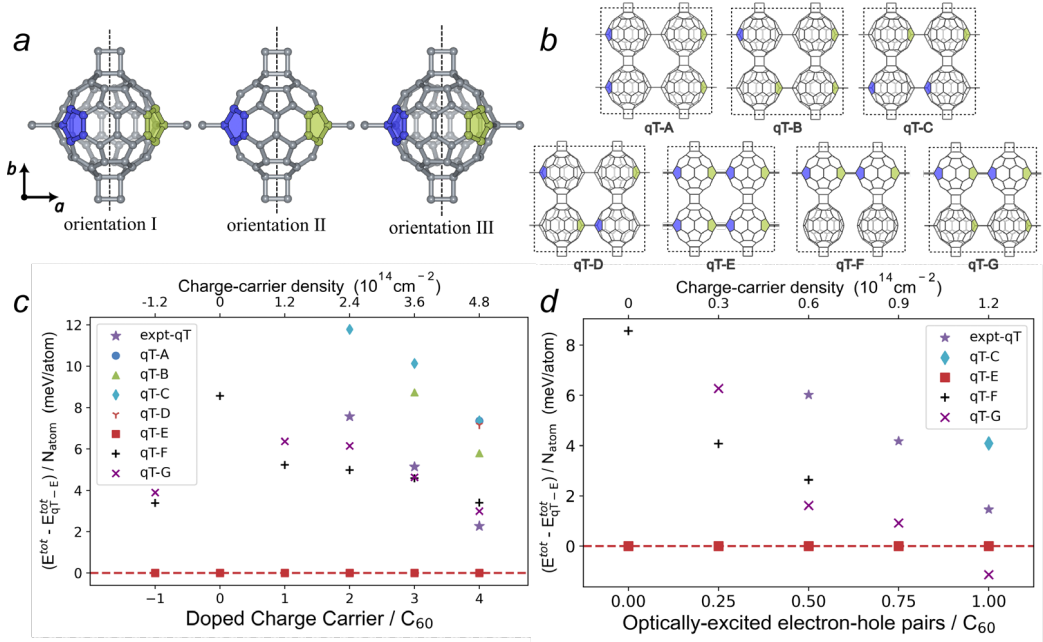

Fig. 3 (a) Three orientations of C$_{60}$ units in quasi-tetragonal (qT) structures. In orientation I, the C atom at a vertex of the yellow pentagon is bonded to a neighboring C$_{60}$; in orientation II, both the yellow and blue pentagons have a carbon atom bonded with the corresponding adjacent C$_{60}$ clusters; and in orientation III, the C atom at a vertex of the blue pentagon is bonded to the adjacent C$_{60}$. (b)The top views of qT-A, qT-B, qT-C, qT-D, qT-E, qT-F, and qT-G structures. (c) Relative energies per atom of qT structures at doping conditions. (d) Relative energies of quasi-tetragonal polymorphs when 0, 0.25, 0.5, 0.75 and 1 electron-hole pairs per C$_{60}$ are excited. Here, we only plot the data points that are converged in DFT structural optimization. The energy of the qT-E structure is used as reference.

Previously, several experiments reported light-induced phenomena of C$_{60}$-based materials, such as photo-induced polymerization of C$_{60}$[42-43] and photo-induced superconductivity in K$_3$C$_{60}$[44]. The significant impact of electrostatic doping on qT polymorphs and strong light-matter coupling in ultrathin materials motivate us to explore the effects of optically excited electron-hole pairs on qT polymorphs. Here we use delta self-consistent field method to simulate effects of depopulating the highest occupied bands and populating the lowest unoccupied bands[45-46]. By comparing the energies of qT structures, we found the relative energies of qT polymorphs also sensitively depend on the concentration of photo-excited electrons and holes, as shown in Fig. 3 (d). For instance, the expt-qT structure becomes stable and its energy relative to the qT-E structure decreases as the number of excited electrons per $C_{60}$ increases. When less than 1 electron hole pair per $C_{60}$, qT-E is the ground state. When 1 electron-hole pairs per $C_{60}$ (i.e., $1.2 \times 10^{14}$ cm$^{-2}$) are excited, qT-G becomes approximately 2 meV/atom lower than the qT-E structure. In other words, the qT-G structure that is unstable in the undoped condition may be stabilized when optically excited electron-hole concentration reaches around $1 \times 10^{14}$ cm$^{-2}$, which is on the same order of magnitude as that experimentally achieved in monolayer transition metal dichalcogenides[47,48].

Different from few-layer expt-qT, the experimental quasi-hexagonal (qH) phase (abbreviated as expt-qH) is synthesized from $Mg_4C_{60}$, where each $C_{60}$ unit is covalently bonded with six neighboring clusters. The $C_{60}$ clusters have two orientations in the expt-qH phase, which can be distinguished by examining the pentagon on the top of $C_{60}$, as highlighted in Fig. 4 (a). The $C_{60}$ clusters bonded via [2+2] cycloaddition bonds along the $\vec{a}$-axis have the same orientation, while $C_{60}$ clusters bonded via C-C single bonds along $\vec{b} + \vec{a}$ or $\vec{b} - \vec{a}$ directions are oriented in another direction. Similar to the tetragonal phases, one can construct various quasi-hexagonal structures by combining these two $C_{60}$ orientations. In addition to the expt-qT structure, we considered three other quasi-hexagonal monolayer $C_{60}$ polymers, denoted as qH-A, qH-B, and qH-C phases. Fig. 4 (b) schematically displays these four qH structures and highlights the top pentagons to better distinguish them. Fig. 4 (c) shows the phonon dispersions of monolayer qH-A, B, and C structures, which are all free of imaginary modes signifying their dynamic stability. The band structures of four qH polymorphs are presented in Supplementary Fig. S5, showing band gaps in range from 0.44 eV to 0.70 eV from PBE calculations. Therefore, one can distinguish these qH polymorphs experimentally by measuring band gaps.

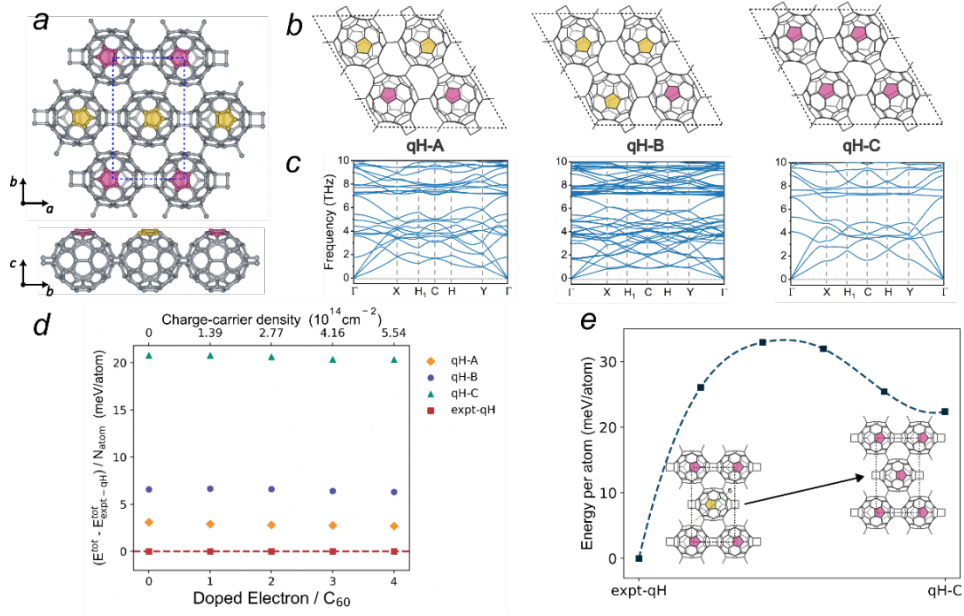

Fig. 4 (a) The top and side views of experimental quasi-hexagonal (qH) structures. The top pentagons are highlighted in colors to illustrate two different $C_{60}$ orientations. (b) Several quasi-hexagonal polymorphs constructed by combining two orientations of $C_{60}$ clusters. These qH polymorphs are denoted as qH-A, qH-B, and qH-C. (c) Phonon dispersions of qH polymorphs under undoped conditions. For clarity, only phonon modes with frequencies lower than 10 THz are shown. (d) Relative energies per atom of qH-A, qH-B, qH-C , and expt-qH. The expt-qH structure is set to be the reference. When doped with 1 e$^-$ per $C_{60}$, the corresponding doped carrier density is about $1.39\times 10^{14}$ cm$^{-2}$. (e) The changes in energy along the transition from expt-qH to qH-C structure.

By comparing the energies of qH structures upon electrostatic doping of electrons (Fig. 4(d)), we find the relative energies of four qH structures show negligible changes, which is in stark contrast to the large doping-induced energy changes found in quasi-tetragonal structures. Among them, the expt-qH structure remains the ground state when the doping carrier concentration is between 0 to 4 e$^-$/$C_{60}$. The energies of qH-A and qH-C polymorphs are close and about 3 meV/atom above that of expt-qH structure, while the energies of the qH-B and qH-C structures are 7 and 21 meV/atom higher than the expt-qH structure, respectively. Using the nudged elastic band (NEB) method, we calculate the energy barrier for rotating a $C_{60}$ cluster in the expt-qH structure. As shown in Fig. 4 (e), the energy barrier for $C_{60}$ rotation in the expt-qH structure is more than 30 meV/atom, much higher than that in quasi-tetragonal structures. The higher energy barrier can be qualitatively explained by the fact that rotating $C_{60}$ in the expt-qH structure breaks more inter-cluster bonds than the cases of qT structures.

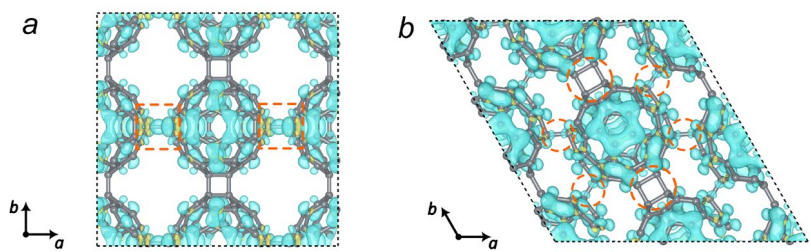

Fig. 5 (a) The charge density difference between the undoped state and that doped with 4 electrons per $C_{60}$ for expt-qT. (b) The charge density difference between undoped state and the state doped with 4 electrons per $C_{60}$ for expt-qH. The cyan and yellow isosurfaces correspond to charge density changes of 0.00105 e Bohr$^{-3}$ and -0.00105 e Bohr$^{-3}$, respectively.

Our calculations reveal the relative energies of qT structures strongly depend on the concentration of doped carriers, whereas the energies of qH structures show weak dependence on doping condition. To better understand the distinctive doping effects on qH and qT polymorphs, we analyzed the changes of charge density distribution between the undoped condition and 4 e$^-$/$C_{60}$ doping condition. As highlighted by the dashed lines in Fig. 5 (a), the charge density for doped electron substantially accumulates at the C-C single bond along the $a$-direction of the expt-qT structure when it is doped with electrons. Such charge accumulation strengthens the C-C single bonds and thereby enhances the stability of the expt-qT structure. By examining the charge density changes of other qT structures, we find electron doping has mixed effects on the C-C bonds along the $a$-direction in different qT polymorphs (as shown in Supplemental Figure S3), resulting in evident doping-induced changes of their relative energies. Different from qT polymorphs, electron doping introduces small accumulation of charge density at the inter-$C_{60}$ bonds of all qH structures (as shown in Supplemental Figure S4). As highlighted by the dashed lines in Fig. 5(b), doped electrons contribute little near the inter-cluster bonds in the expt-qH structure. In other words, electrostatic doping has minor effects on the inter-cluster covalent bonds in qH $C_{60}$ polymers, and hence leads to a limited impact on the relative energies of qH polymorphs.

## CONCLUSIONS

In summary, our finding enriches the family of fullerene-assembled 2D materials that are structurally related by cluster rotation. We show that electrostatic and optically-excited electron-hole pairs can effectively tune the relative energies of various qT phases. In particular, the monolayer phase of expt-qT, which is unstable at undoped condition, can be stabilized by electrostatically doping with electrons. The energy barrier of rotating $C_{60}$ clusters in monolayer qT polymorphs is lower than that of quasi-hexagonal phases, suggesting a mixture of several qT structures or the orientation-glass state with random $C_{60}$ orientation. Several quasi-hexagonal polymorphs with dynamic stability and energies close to those of expt-qH are also proposed. Different from qT structures, the relative energies between qH structures cannot be effectively changed by electron doping.

## COMPUTATION METHODS

DFT calculations were performed using the plane-wave basis and the projector augmented wave approach as implemented by the Vienna ab initio simulation package (VASP 5.4.4) [49]. We used the Perdew-Burke-Ernzerhof (PBE) functional[50] and the D3 correction[51] to treat exchange-correlation effects and dispersion interaction, respectively. The configurations of valence electrons considered in PAW potentials for C is $2s^22p^2$. A plane-wave cutoff of 400 eV, and a 3×3×1 Monkhorst−Pack k-point mesh for both qT and qH structures were used[52]. To model charged monolayer $C_{60}$, we set the out-of-plane lattice parameter $c$ to 75 Å, which corresponds to the vacuum layer thickness of ~ 68 Å. Tests of convergence with respect to lattice parameter $c$ are presented in Fig. S6 and discussions in the Supplementary Information. All structures were optimized until the energy differences between consecutive ionic steps are less than $10^{-4}$ eV/supercell and the force felt by each atom is less than 0.01 eV/Å. Phonon dispersions were calculated in the PHONOPY[53] package via the frozen-phonon method. VESTA[54] package is used for the illustration of atomic structures. We performed nudged elastic band [39-40] calculations using VTST (Transition State Tools for VASP).

The NEB calculations were performed with a convergence criterion of 0.03 eV/Å, and the energy profile was obtained by interpolating the energies of the intermediate states.

## ACKNOWLEDGMENTS

This work is supported by the National Natural Science Foundation of China (12104080, 91961204) and the Fundamental Research Funds for the Central Universities (DUT22LK04, DUT22ZD103). The authors acknowledge the computer resources provided by the Supercomputing Center of Dalian University of Technology, Shanghai Supercomputer Center, and Sugon supercomputer centers.

# Supplementary materials for

# Rich structural polymorphism of monolayer $C_{60}$ from cluster rotation

Xueao Li[1#], Fan Zhang[1#], Xuefei Wang[1], Weiwei Gao[1*], and Jijun Zhao[1]

1. Key Laboratory of Material Modification by Laser, Ion and Electron Beams (Dalian University of Technology), Ministry of Education, Dalian 116024, China

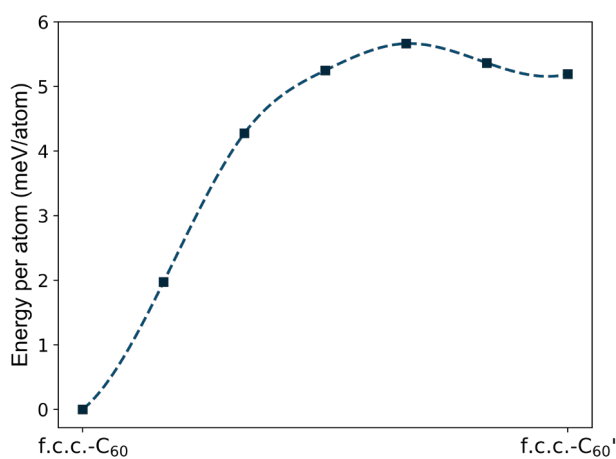

Figure S1. The changes in energy along the transition path of rotating a $C_{60}$ in f.c.c-$C_{60}$ crystal. Here f.c.c-$C_{60}$' represents the state where one of the $C_{60}$ clusters change its orientation, as depicted in Figure S2. A conventional supercell with four $C_{60}$ clusters is used to simulate the rotation process. The rotating energy barrier between the two states is estimated to be about 6 meV/atom.

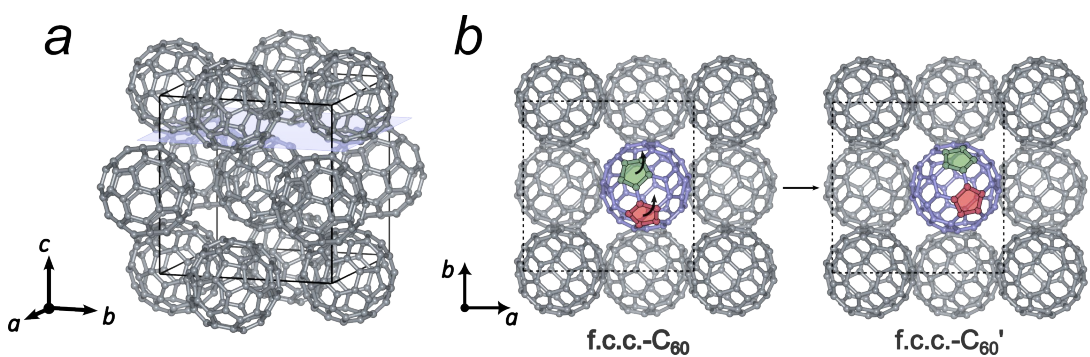

Figure S2. (a) Illustration of face-centered cubic (f.c.c.) $C_{60}$ crystal structure. (b) Top view along the crystal surface shown in (a). The rotation process from f.c.c.-$C_{60}$ to f.c.c.-$C_{60}$' shown in Figure S1. The highlighted pentagons illustrate the orientation of the rotating $C_{60}$.

---

* Corresponding Authors. Email: weiweigao@dlut.edu.cn

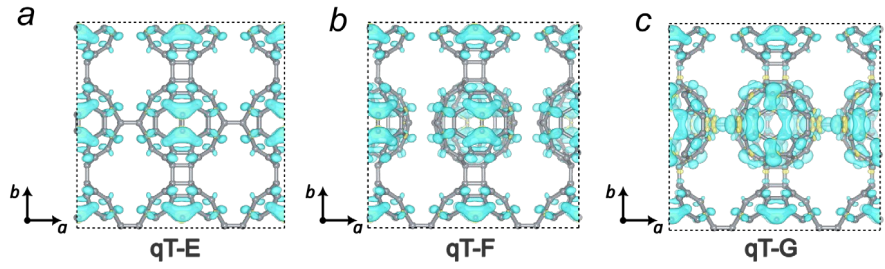

Figure S3(a-c) The top view of the charge density difference between the undoped state and 4 doped electrons per $C_{60}$ for qT-E, qT-F, and qT-G. The cyan and yellow iso-surfaces correspond to charge density changes of 0.00105 e Bohr$^{-3}$ and -0.00105 e Bohr$^{-3}$, respectively. Note the doped electrons do not accumulate at the inter-cluster bonds in qT-E and qT-F structures, but contributed significantly to the charge at the inter-cluster bonds in the qT-G structure.

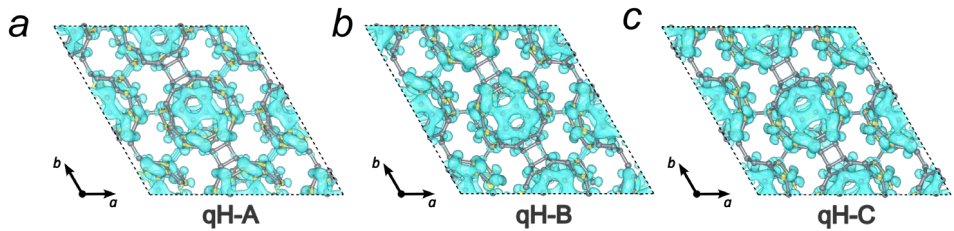

Figure S4 (a-d) The top view of the charge density difference between the neutral state and 4 doped electrons per $C_{60}$ for qH-A, qH-B, qH-C, and qH-D. The cyan and yellow iso-surfaces correspond to charge density changes of 0.00105 e Bohr$^{-3}$ and -0.00105 e Bohr$^{-3}$, respectively. For these three qH structures, the doped electrons do not contribute significantly to the inter-cluster bonds.

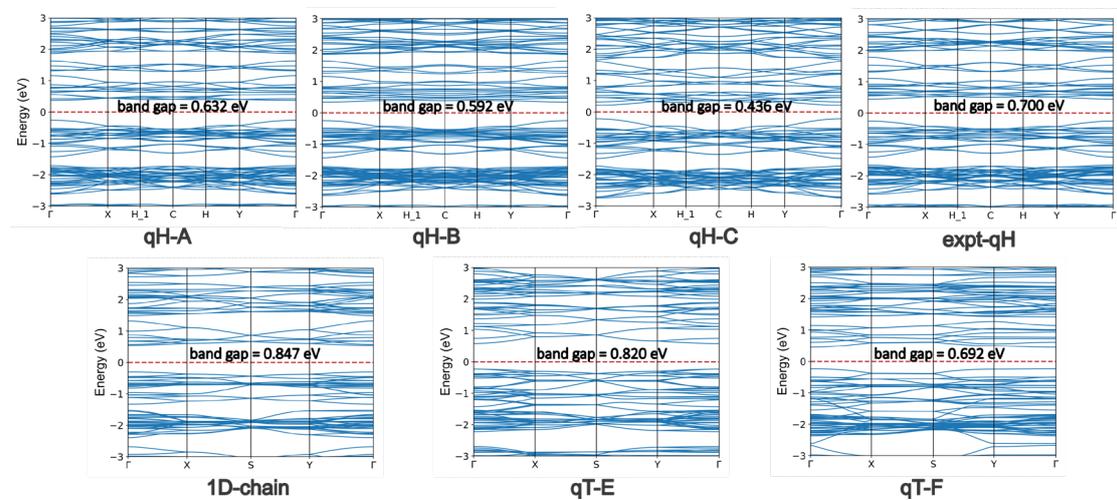

Figure S5. The band structure of monolayer $C_{60}$ polymorphs, namely, qH-A, qH-B, qH-C, expt-qH, 1D-chain, qT-E, and qT-F.

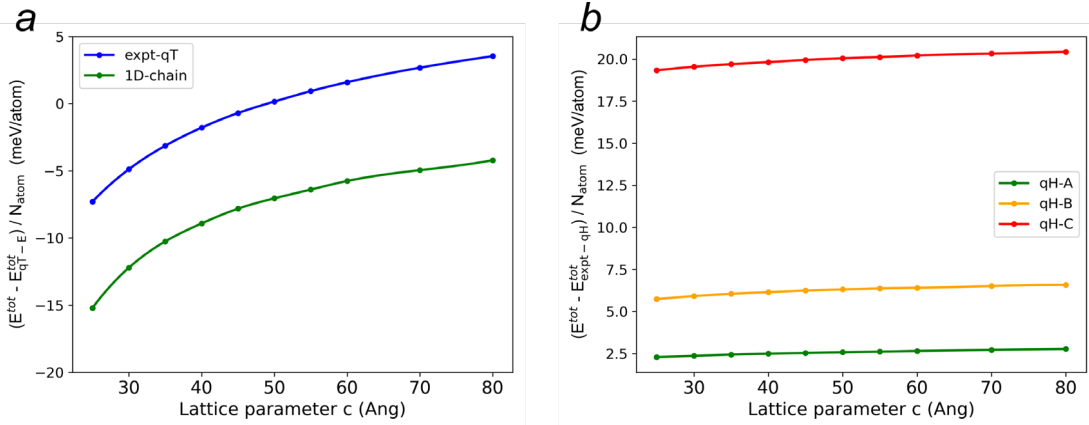

Figure S6. (a) The energies of expt-qT and 1D-chain structures doped with 4 electrons per $C_{60}$ relative to that of the qT-E structure calculated with different vacuum space. (b) The energies per atom of qH-A, B, C structures doped with 4 electrons per $C_{60}$ relative to that of the expt-qH structure calculated with different vacuum space. The thickness of vacuum space is tuned by adjusting the out-of-plane lattice parameter $c$ of the supercell.

As implemented in many plane-wave based density functional theory codes, such as VASP, we simulated a charged monolayer $C_{60}$ by using a jellium background to compensate the extra charges added to the system. It is a well-known challenge to simulate a charged slab system, as the energy converges slowly with the thickness of vacuum space when it is modeled in a supercell. Since we compare the relative energies, the spurious interactions between periodic images and the jellium background largely cancels out as we take the differences between the energies of similar structures (like quasi-tetragonal or quasi-hexagonal polymorphs). Figure S6 shows the convergence of relative energies with the thickness of vacuum space padded along the out-of-plane direction. As illustrated in Figure S6(a), the energies of qT-expt and 1D-chain structures referenced from the qT-E structure converges gradually as the out-of-plane lattice parameter $c$ increases. The relative energies of qT-expt and 1D-chain by about 0.7 meV/atom as $c$ changes from 70 to 80 Å. Compared to qT structures, the energies of qH-A, B, and C relative to that of the expt-qH structure changes negligibly for $c$ ranging from 30 to 80 Å. For the results presented in the main text, we choose $c = 75$ Å to simulate the charged systems.